\documentclass[10pt]{article}
\makeatletter

\def\eqnarray{%
   \stepcounter{equation}%
   \def\@currentlabel{\p@equation\theequation}%
   \global\@eqnswtrue
   \m@th
   \global\@eqcnt\z@
   \tabskip\@centering
   \let\\\@eqncr
   $$\everycr{}\halign to\displaywidth\bgroup
       \hskip\@centering$\displaystyle\tabskip\z@skip{##}$\@eqnsel
      &\global\@eqcnt\@ne \hfil${{}##{}}$\hfil
      &\global\@eqcnt\tw@
         $\displaystyle{##}$\hfil\tabskip\@centering
      &\global\@eqcnt\thr@@ \hb@xt@\z@\bgroup\hss##\egroup
         \tabskip\z@skip
      \cr
}
\newenvironment{subequations}{%
  \refstepcounter{equation}%
  \mathchardef\c@mainequation\c@equation
  \protected@edef\themainequation{\theequation}%
  \let\theequation\thesubequation
  \global\c@equation\z@
  }{%
  \global\c@equation\c@mainequation
  \global\@ignoretrue
  }
\newcommand{\thesubequation}{\themainequation\alph{equation}}
\newenvironment{subeqnarray}{%
  \subequations
  \@ifnextchar\label{\@lab@subeqnarray}{\eqnarray}
  }{%
  \endeqnarray\endsubequations
  }
\newcommand*{\@lab@subeqnarray}[2]{#1{#2}\eqnarray}
\@addtoreset{equation}{section}
\def\theequation{\@arabic\c@section.\@arabic\c@equation}
\makeatother

\usepackage{graphics}
\def\d{\,{\rm d}}
\def\e{{\rm e}}
\def\sh{\mathop{\rm sh}\nolimits}
\def\F{F^2_{\varphi z}}
 
\newcommand{\dil}[1]{\phi_{,\,{#1}}^2} 
\newcommand{\be}[1]{{{\bf e}}_{\hat{#1}}}
\newcommand{\eref}[1]{(\ref{#1})}
\def\sref{\ref}
\def\fl{\noindent}
\def\case{\frac}
\def\Or{{\rm O}}
\def\fref{figure~\ref}
\def\tref{table~\ref}

\makeatletter
\def\section{\@startsection {section}{1}{\z@}%
  {-3.5ex \@plus -1ex \@minus -.2ex}%
  {2.3ex \@plus.2ex}%
  {\null\hfil\normalfont\bfseries}}
\makeatother

\begin{document}

\title{Charged perfect fluid and scalar field coupled to gravity}

\begin{center}{
{\bf CHARGED PERFECT FLUID AND SCALAR FIELD 
COUPLED TO GRAVITY}\\[3mm]
P. KLEP\'{A}\v{C}\footnote
{E-mail: {\tt klepac@physics.muni.cz}} 
and J. HORSK\'Y\footnote{E-mail: 
{\tt horsky@physics.muni.cz}}\\[2mm]
{\it Institute of Theoretical Physics and 
Astrophysics, Faculty of Science,\\ Masaryk 
University, Kotl\'{a}\v{r}sk\'{a} 2, 611 37  Brno, 
Czech Republic}}\end{center} 

{\small Charged perfect fluid with vanishing 
Lorentz force and massless scalar field is studied 
in the case of stationary cylindrically 
symmetric spacetime. The scalar field can depend 
both on radial and longitudinal coordinates. 
Solutions are found and classified according to scalar field 
gradient and magnetic field relationship. Their 
physical and geometrical properties are examined 
and discussion of particular cases, directly 
generalizing G\"{o}del-type spacetimes, is presented.\\

\noindent PACS numbers: 04.20 Jb, 04.20 Gz}

\section{Introduction}

Since G\"{o}del original work \cite{Godel} in 1949 
the rotating charged perfect fluid cosmology has 
become an attractive subject for many scientists. 
These cosmological models, mostly restricted to 
Lorentz force-free cases, are remarkable as they 
exhibit the interplay of rotation, gravitation 
and electromagnetism. Wright \cite{Wright} 
constructed inhomogeneous class of electrically 
neutral perfect fluid solutions. Banerjee and 
Banerji \cite{Banerji} gave electromagnetic 
G\"odel-type solution, considering longitudinal 
magnetic field parallel to rotation axis. Solution 
proposed by Som and Raychaudhuri \cite{Som} is 
electromagnetic analog of van Stockum solution 
\cite{Kramer}. Bonnor \cite{Bonnor} obtained 
solutions for axially symmetric dust spacetimes 
both for non-vanishing and vanishing Lorentz force, 
the latter being general one (if rotation is rigid). 
In ninetieth Mitski\'evi\v{c} and Tsalakou 
\cite{Mit} found inhomogeneous charged $G_3$ 
spacetimes using Horsk\'y-Mitskievi\v{c} conjecture 
\cite{HM} and their results were generalized 
to an arbitrary magnetic 
potential by Upornikov \cite{Upornikov}. Physical 
interpretation of Lorentz force-free charged fluids 
in general relativity can be found in \cite{MitMex}. 
Recently, Klep\'a\v{c} and Horsk\'y submitted $G_3$ 
spacetimes for generally non-vanishing Lorentz force 
\cite{Klep}. 

The reason why to study the coupling of the 
charged perfect fluid and the scalar field 
to gravity relies on the fact that in modern 
unified theories such as string and 
superstring theories (\cite{{Barrow},{Kanti}}), there appear 
effective actions reproducing, at the 1-loop level, 
the Einstein field equations enriched in four 
dimensions by a contribution of one or  more scalar 
fields (dilaton, axion, etc.). Therefore it is natural, 
as a first step, to search for solutions of the 
Einstein equations describing  the coupling of a 
charged perfect fluid and a dilaton to gravity. 
Then the fluid can be viewed as an very rough 
approximation of the fermionic matter in the theory.

The paper is organized as follows. In section 
\ref{sec2} basic assumptions and equations are 
introduced and in section \ref{sec3} main 
solutions are derived. Section \ref{sec4} 
specializes general results of section 
\ref{sec3}, intending to clarify their 
connection with G\"odel-type metrics. There 
is the remark on scalar field behaviour and 
brief conclusion in \sref{conclusion}.                

\section{The Einstein-Maxwell equations\label{sec2}}
We are searching for stationary cylindrically symmetric 
cosmological models with charged perfect fluid and 
scalar field coupled to gravity. Let us introduce 
locally Lorentz coframe
$\d s^2=\eta_{\alpha\beta}\Theta^{\hat{\alpha}}\otimes
\Theta^{\hat{\beta}}$
with basis 1-forms 
\begin{equation}\label{basis}
\begin{array}{ll}
\Theta^{\hat{0}} = \d t+f\d\varphi\ ,  \qquad & 
\Theta^{\hat{1}} = l\d\varphi\ , \\
\Theta^{\hat{2}} = \e^\gamma\d z\ , \qquad \qquad &  
\Theta^{\hat{3}} = \e^\delta\d r\ ,
\end{array}
\end{equation}
where $(t, \varphi,z,r)=(x^0,x^1,x^2,x^3)$ are 
local cylindrical coordinates on a manifold, 
{\it adapted} to the Killing vector fields 
$\partial_t,\ \partial_\varphi,\ \partial_z$, 
and $f,\ l,\ \gamma,\ \delta$ are functions of 
$r$ alone. If one assumes that a spacetime we are looking 
for is filled with the perfect fluid of pressure 
$p(r)$ and energy density $\mu(r)$ moving in its 
rest frame with velocity $u=\Theta^{\hat{0}}$, 
then the motion of the fluid is characterized by 
geodesic rigid rotation around $z$ axis, $\dot{u}=0,\ 
\Theta = \sigma =0.$ Thus velocity is the Killing 
vector field. 

Since in this paper we concentrate on 
cases with vanishing Lorentz force, in the 
fluid comoving frame there is purely magnetic 
field present. Excluding cylindrically 
symmetric spacetimes with currents parallel 
to $z$-axis, due to symmetry only 
\begin{equation}\label{magnetic}
B_{\hat{r}}=F_{\varphi z}(r)l^{-1}\e ^{-\gamma}\ ,\qquad
B_{\hat{z}}=F_{r \varphi}(r)l^{-1}\e ^{-\delta}
\end{equation}
components of the magnetic field survive.
Note that here we consider only $r$ dependent 
electromagnetic field. In order to satisfy 
condition $\d F=0$ it follows that 
$F_{\varphi z}$ is constant. If fluid particles 
are charge carriers, the Maxwell equations with 
sources $-*\d *F=4\pi j=4\pi \rho u$ 
give expressions for $z$-component of 
the magnetic field and for the invariant charge 
density $\rho .$ Denoting for future convenience 
$M=l\e^{-\gamma+\delta}$ one has  
\begin{equation}\label{constraint}
F_{r\varphi}=Bl\e ^{-\gamma+\delta}\equiv BM\ ,
\qquad 4\pi\rho =-\frac BM\frac{\d f}{\d r}\ \e^{-2\gamma}\ , 
\end{equation}      
$B$ being constant characterizing longitudinal 
component of the magnetic field.

Finally we incorporate massless scalar 
field $\phi$ into the theory. Owing to 
the symmetry of the problem, $\phi$ is, 
in general, assumed to be function of 
$r$ as well as of $z.$ Stress-energy 
tensor associated with the scalar field  
is taken in the usual way
\begin{equation}\label{Tscalar}
8\pi T_{\rm scal}=\left[\left(\be {\mu}\phi\right)\ 
\left(\be {\nu}\phi\right) 
-\frac 12 \eta_{\mu\nu}\left(\nabla \phi\right)^2\right]
\Theta^{\hat{\mu}}\otimes\Theta^{\hat{\nu}}\ ,
\end{equation}
where $\be {\mu}$ are basis vector fields dual to basis 1-forms 
\eref{basis}
\begin{equation}
\begin{array}{ll}
\label{basisvec}
\be {0}=\partial_t\ ,\qquad \qquad & \be 1=l^{-1}\left(
\partial_\varphi-f\partial_t \right)\ ,\\
\be 2=\e ^{-\gamma}\partial_z\ ,\ \qquad & 
\be 3=\e^{-\delta}\partial_r\ .
\end{array}
\end{equation}
\newcommand{\der}[1]{{\frac{\d {#1}}{\d r}}}
\newcommand{\pder}[2]{\frac{\partial {#1}}{\partial {#2}}}
\def\q{\qquad}
Appropriate Einstein-Maxwell equations 
read (with $c=G=1$) 
\def\ZZZ#1#2{\hbox to\hsize{$\displaystyle
  \hbox to3.8cm{\hss$\displaystyle#1{}$}#2$\hss}}
\begin{subeqnarray} 
\ZZZ{\frac{\d}{\d r}\left[\frac 1M\der f\right]=}{0\ ,}
  \label{omega} \\[2mm]
\ZZZ{\frac{\d}{\d r}\left[\frac 1M\der \gamma\right]=}
  {-\frac 1M\left(\pder{\phi}{r}\right)^{\!\!2}
  +\frac{2M}{l^4}\F \ ,}
  \label{ff-rr} \\[2mm]
\ZZZ{\frac M2\frac{\d}{\d r}\left[\frac 1M \frac{\d l^2}{\d r}\right]=}
  {16\pi  l^2p \, \e^{2\delta}\ ,}
  \label{zz+rr}\\[2mm]
\ZZZ{\frac{\d}{\d  r}\!\left[\frac{\e^{-2\gamma}}{M}
  \frac{\d}{\d r}\left(l^2\e^{2\gamma}\right)\right]=}
  {\frac{16\pi(p-\mu)}{M}l^2\e^{2\delta}
  +\frac 1M\left(\der f\right)^{\!\!2}
  -2M\left(\pder {\phi}{z}\right)^{\!\!2}}
  \nonumber\\
\ZZZ{}{-4\F l^{-1}\e^{-\gamma+\delta}\ ,}
  \label{tt-rr} \\[2mm] 
\ZZZ{\der\gamma \frac{\d \, l^2}{\d r}=}
  {16\pi l^2p \, \e^{2\delta}-\frac12 
  \left(\der f\right)^{\!\!2}+2B^2M^2-2\F
  \e^{2\delta-2\gamma}}
  \nonumber\\
\ZZZ{}{+l^2\left(\pder {\phi}r\right)^{\!\!2}
  -M^2\left(\pder {\phi}r\right)^{\!\!2}\ ,}
  \label{intBianchi}\\[2mm]
\ZZZ{\pder\phi r\pder\phi z=}{2BMl^{-2}F_{\varphi z}\ ,}
  \label{zr}
\end{subeqnarray}
with $\phi$ subject to scalar equation of motion
\begin{equation}
\frac 1M\frac\partial{\partial r}\left(\frac
{l^2}{M}\pder\phi r\right)+\frac{\partial^2 \phi}
{\partial z^2}=0\, .
\label{scalmotion}
\end{equation}

Equation \eref{intBianchi} is integral 
of the Bianchi identity that claims constant 
pressure $p$ (by virtue of geodesic motion) 
if \eref{scalmotion} holds, which in turn 
appears to be satisfied identically if the 
Einstein equations are. Therefore after 
eliminating $p$ from remaining Einstein 
equations one has altogether seven independent 
equations \eref{omega}-\eref{tt-rr}, \eref{zr} 
and two non-trivial Maxwell equations for eight 
unknown functions $f,\ l,\ \gamma,\ \delta,
\ \mu,\ \rho,\ F_{\varphi r}$ and $\phi$.  
One degree of freedom corresponds to the 
possibility of choosing arbitrarily the scale 
function $\delta$.
\renewcommand{\dil}[1]{\phi_{\,{#1}}^2} 
Equation \eref{omega} gives $f=2\Omega m+D$, 
where $m=\int M\d r$. In case of no radial 
magnetic field $m$ represents (non-uniquely) 
angular component of magnetic potential (i.e. 
vector potential is $A=Bm\d \varphi$). Constant 
$\Omega$~ is the rate of the rigid rotation of 
matter around $z$-axis, with vorticity covector 
$\omega=\frac 12 *\left(u\wedge\d u\right)=
\frac 12\frac {f'}{M}\d z=\Omega \d z$, 
with prime denoting derivative with respect to $r$. 
Unimportant constant $D$ will be omitted further. 

Because of the independence of $r$ and $z$-coordinates 
it follows from \eref{tt-rr}, \eref{ff-rr} and \eref{zr} 
that $\phi$ is expressed like 
\begin{equation}\label{dilaton}
\phi = \phi_0+\phi_1z+\phi_2\int\frac{M}{l^2}\d r\,, 
\qquad \phi_2=\frac{2BF_{\varphi z}}{\phi_1}\ ,
\end{equation}
where $\phi_0,\ \phi_1$ are constants and the latter 
equality being valid provided that $\phi_1\neq 0$.

By integrating of combination of \eref{zz+rr}, 
\eref{intBianchi} and \eref{ff-rr}, and excluding 
$\gamma$, one obtains a non-linear equation 
relating $l^2$ to $M$ 
\begin{equation}\label{nonlin}
\frac {l^4}M\frac{\d }{\d r}\left[\left(
\case 1M\case {\d l^2}{\d r}+
\left(4B^2-2\phi^2_1- 4\Omega^2\right)m+4k
\right)l^{-2}\right]=
4\F -2\phi^2_2\ .
\end{equation}
In \eref{nonlin} an integration constant is 
denoted $4k$ in conformity with \cite{Klep}. 

We proceed further by splitting solutions into two groups 
according to whether right-hand side of \eref{nonlin} 
vanishes or not. 

\section{Exact solutions\label{sec3}}

\subsubsection*{Case I: $\dil 2=2\F $.\label{caseI}} 
Radial part of scalar field gradient balances 
the radial component of magnetic field. This 
case will be itemized in two subcases.

(a) If $F_{z\varphi}\neq 0$ then solution of the 
Einstein-Maxwell system reads ($P=8\pi p$)
\begin{equation} 
\fl \d s^2\!=\!\left(\!\d t\!+\frac {2\Omega}C 
\gamma\d\varphi\right)^2\!\!
-\!\left(\!\frac {P}{C^2}\e^{2\gamma}\!
-\!\frac{2\Omega^2}{C^2}\gamma
\!+\!\nu\right)\!\!\d\varphi^2\!-\e^{2\gamma}\d z^2 
\! -\frac{\gamma^{\prime 2}\e^{2\gamma}\d r^2}
{P\e^{2\gamma}\!-\!2\Omega^2\gamma\!+\!C^2\nu} 
\label{Wright}\ , 
\end{equation}
with $C$, $\nu$ as integration constants and $\gamma$ an 
arbitrary non-constant twice differentiable function of 
$r$. Metric \eref{Wright} is that given by Wright more 
then thirty years ago \cite{Wright}. In his work, Wright 
considered a dust spacetime with non-vanishing cosmological 
constant. 

Thus we have succeeded in finding an alternative source 
for the Wright metric. In this way ambiguity in sources for 
the G\"odel metric disscussed in \cite{ReboucasI} is now 
generalized to ambiguity in sources for the Wright metric. 
Here we have come across with non-zero charge and energy 
densities given by  
\begin{equation}\label{Wrightdens}
2\pi\rho = -B\Omega\e^{-2\gamma}\ ,\ \mu =
\frac{1}{4\pi}\left(2\Omega^2-B^2-l^{-2}\F \right)
\e^{-2\gamma}-3p\ ,
\end{equation} 
and scalar field $\phi$, that is defined by \eref{dilaton}, 
and cannot be determinated explicitly. In second equation 
of \eref{Wrightdens} the rotation, represented by vorticity 
scalar $\omega^2=|\omega_{\nu}\omega^{\nu}|=\Omega^2
\e^{-2\gamma}$, is compensated by addition of magnetic 
energy density and ``specific'' mass density $\mu+3p$ 
(it is interesting to compare \eref{Wrightdens} 
with equation (6.2) in Bonnor \cite{Bonnor}). 

The integration constants in \eref{Wright} are to be chosen 
properly in order to ensure Lorentzian signature, energy 
conditions and even regularity at the origin, if desired 
\cite{Santos}. In order to fulfil the energy conditions, 
$\omega^2 \geq P$ or $2\omega^2 \geq P$ depending on whether 
the pressure is positive or negative, respectively.

The qualitative decription of the radial 
magnetic and scalar fields 
behaviour in the vicinity of the rotation axis 
$r=0$ can be carried out as follows. 
Equation \eref{Wright} restricted to the 
$(r,\varphi)$ subspace for small values of $r$ reads
\begin{equation}\label{rphi}
\d s_2^2 =C^{-2}l^{-2}(\d\e^{\gamma})^2 + l^2\d\varphi^2\ ,
\end{equation}
and is regular for $l =0$ provided $\e^{\gamma} \sim 
K \pm Cl^2/2$, $K$ is constant (see exact values 
of the integration constants given below). Then 
from \eref{dilaton} one has 
\begin{equation}\label{sing}
\phi = \phi_0+\phi_1z+\phi_2C^{-1}\int l^{-2}\d\gamma =
\phi_0+\phi_1z\pm \phi_2K^{-1}\ln \frac{2l}
{\left|Cl^2\pm 2K\right|^{1/2}}\ .
\end{equation}
In this way the radial part scalar field behaves as 
$\phi \sim \phi_0+\phi_1z\pm \frac{\phi_2}K\ln l$, 
which is the singular solution of the Laplace equation 
of motion for line of a scalar charge at $l~=~0$. 
Divergences in $z$-infinities represent additional 
sources of scalar charge and they are common 
for scalar fields in G\"odel-type metrics both 
in classical \cite{ReboucasI} as well as 
in the string theories \cite{{Barrow},{Kanti}}
(solution \eref{shdilaton} below is singular 
for the same reason). 
For small $r$ the scalar field 
gradient radial part goes as $\pder\phi r\sim r^{-1}$. 
On the other hand the radially
pointing magnetic field $B_{\hat{r}}$ defined
in \eref{magnetic} is also singular on the $l=0$ line, 
that acts like a linear source of the 
radial magnetic field (which is the 
same as in the classical theory), and can 
be interpreted as a ``line of magnetic charge''. 
In the neighbourhood of the rotation axis also
$B_{\hat r}\sim r^{-1}$ and divergences produced 
by scalar field gradient along with 
that produced by radial magnetic field are canceled 
mutually keeping the total stress-energy tensor regular 
on the rotation axis.

(b) Subcase $F_{z\varphi}= \phi_2=0$ yields metric
\begin{eqnarray} 
\d s^2 \!&=&\!\left(\!\d t\!+\!\frac {2\Omega}C
\gamma \d \varphi\right)^2
\!-\!\left(\!\frac{P}{C^2}\e^{2\gamma}\!+\!\frac{\lambda}{C}\gamma
+\nu\right)\!\d\varphi^2\!-\e^{2\gamma}\d z^2 
-\frac{\gamma^{\prime 2}\e^{2\gamma}\d r^2}
{P\e^{2\gamma}+C\lambda\, \gamma+C^2\nu}\ ,\nonumber\\
\lambda &=&\frac{2B^2-2\Omega^2-\dil 1}{C}\ ,\ 
\nu=\frac{\lambda+4k}{2C}\ .\label{genscalar}
\end{eqnarray}

The charge density, pressure and the energy density are 
obtained from \eref{Wrightdens} for $F_{z\varphi}=0$. 
Scalar field becomes equal to $\phi=\phi_0+\phi_1z$ and 
does not enter the energy density, which could be expected 
because of the lack of interaction between charged fluid 
and scalar field. If both $\phi_1$ and $\phi_2$ are zero, 
\eref{genscalar} convertes to Lorentz force-free solution 
presented in \cite{{Klep},{Upornikov}}.  

If one requires \eref{Wright} or \eref{genscalar} to be 
cylindrically symmetric and regular at the origin, one 
has to impose axial symmetry condition
\begin{equation}\label{axsym}
X\equiv\partial_\varphi\cdot\partial_\varphi
=g_{\varphi\varphi} \propto \Or (r^2)
\end{equation}
as $r\rightarrow +0$, 
and elementary flatness condition
\begin{equation}\label{regularity}
\frac{X_{,\alpha} X_{,\beta} g^{\alpha\beta}}
{4X}=\frac{g^2_{\varphi\varphi ,r}}{4g_{\varphi\varphi}g_{rr}}
\rightarrow 1\ 
\end{equation}
when $r\rightarrow +0$ (\cite{{Kramer},{Santos}}). 
Assuming for simplicity $\gamma \propto r^2$, when $r$ tends 
to zero (in practise $\gamma \propto \case C2r^2$), 
the conditions \eref{axsym} and \eref{regularity} 
applied on \eref{genscalar} show following relations
\begin{equation}\label{con}
\case 12C\lambda +P= C\ ,\ C^2\nu +P=0\ .
\end{equation}

The metric \eref{genscalar} turns out to be algebraically 
general, except for the hypersurfaces (which could be of even 
infinite number for given integration constants, e.g. for 
periodic~$\gamma$) on which
\begin{equation}\label{Petrgenscal}
\left|4\Omega^2+C\lambda\right| =4\left|\Omega Cl\right|\ ,
\end{equation}  
where it belongs to the type II. 

The energy conditions imposed on total stress-energy tensor 
demand inequalities given in the \tref{tab1}. 

\begin{table}[h]
\caption{The energy conditions implied by values of 
$\dil 1$ shown on upper line and $p$ in first column.}

\centerline{
\def\arraystretch{1.4142}
\begin{tabular}{c||c|c}
 &$\dil 1\leq 2B^2$ & $\dil 1\geq 2B^2$ \\
\hline
 $p\leq 0$& $4\Omega^2+\dil 1-2B^2\geq 2P\e^{2\gamma}$&$2\omega^2 \geq P$\\
\hline
 $p\geq 0$& $4\Omega^2+\dil 1-2B^2\geq 4P\e^{2\gamma}$ & $\omega^2\geq P$
\end{tabular}}
\label{tab1} \end{table}

\subsubsection*{Case II: $\dil 2\neq 2\F$.} We take 
non-trivial solution of \eref{nonlin} describing charged 
dust and scalar field distribution. The metric, written 
down in terms of an arbitrary non-constant twice 
differentiable function $m$, becomes 
\newcommand{\La}[1]{\left(\lambda m+{\tilde\nu}\right)^{#1}}
\def\ex{\frac{2\dil 2-4\F}{\lambda^2}}
\begin{equation}
\fl \d s^2=\left(\d t+2\Omega m\d\varphi\right)^2-
\La {}\d\varphi^2 -\La {\sigma}b^{-\sigma}\e^{2Cm}
\left(\d z^2+\frac{m^{\prime 2}\d r^2}
{\lambda m+{\tilde\nu}}\right)\ ,
\label{dust}
\end{equation}
and the charge and energy density are found to be
\begin{equation}
\begin{array}{l}\label{dustdens}
\fl 2\pi\rho\! =\! -\!B\Omega \La{-\sigma}b^{\sigma}\e^{-2Cm}\\[2mm]  
4\pi\mu\! =\!
\left(2\Omega^2-B^2-l^{-2}\F \right)
\La{-\sigma}b^{\sigma}\e^{-2Cm}\ . 
\end{array}
\end{equation}
Constants $\sigma$ and ${\tilde\nu}$ are defined by
\[\sigma=\ex \ ,\ {\tilde\nu}=\nu-\frac \lambda{2C}\sigma\ ,\] 
with constant $b$ being of square length dimension.  

Notice that if $B_{\hat r}$ is everywhere zero the charge 
to mass density ratio has a fixed value through the spacetime. 
This result is consistent with theorem in Bonnor \cite{Bonnor}. 

\noindent The scalar field $\phi$ is given by \eref{dilaton} 
leading to
\begin{equation} \label{dustdilaton}
\e^{\phi}=\La{\frac{\phi_2}{\lambda}}\e^{\phi_0+\phi_1z}\ ,
\end{equation} 
and $b$ is involved in constant $\phi_0$. 

Next we turn to the axial symmetry of \eref{dust}. 
Assuming $m\propto r^2$ as $r\rightarrow 0$, the 
conditions \eref{axsym} and \eref{regularity} read 
\[{\tilde \nu}=0\ ,\qquad \frac{\lambda^2}4
\left(\frac b{{\tilde\nu}}\right)^{\sigma}=1\ ,\]
from which we see that the solution \eref{dust} 
cannot represent cylindrically symmetric spacetime 
regular at the origin unless $\sigma=0$ when it 
coincides with dust version of \eref{Wright}. As a 
matter of fact, there are two possibilities to retain 
\eref{dust} physically admissible. Either relax 
the elementary flatness on the symmetry axis, and  
such spacetimes are sometimes accepted (\cite{Santos}), 
or abandon $\varphi$-coordinate periodicity. 

Putting $m=\frac{\lambda}4r^2,\ {\tilde \nu}=0$ 
and carrying out coordinate transformation 
${\tilde \varphi}=\frac{\lambda}2\varphi$, 
one can establish \eref{dust} in more convenient form 
\begin{equation}
\d s^2=\left(\d t+\Omega r^2\d {\tilde\varphi}\right)^2-
r^2\d {\tilde\varphi}^2 -\left(\frac{\lambda ^2}{4b}\right)^{\sigma}
r^{2\sigma}\e^{\frac{C\lambda}2r^2}
\left(\d z^2+\d r^2\right)\ ,\label{isdust} 
\end{equation}
which strongly resembles Som and Raychaudhuri solution 
\cite{Som}, but differs from it by $g_{zz}$ component, 
which gets more complicated. 

\noindent The spacetime \eref{dust} is again algebraically 
general, except on hypersurfaces
\begin{equation}\label{Petrgenscal}
\left|4\Omega^2+C\lambda+\frac \sigma 2 
\left(\case \lambda l\right)^{\!\!2}\right| = 
\left|\frac{2\Omega l}\lambda \left[2C\lambda+\sigma
\left(\case \lambda l\right)^{\!\!2}\right]\right|\ ,
\end{equation}  
where it is of the type II.  

We finish this section by setting down explicit 
expressions when the energy conditions are satisfied. 
Following inequalities must hold 
\begin{subeqnarray}
2\Omega^2\geq \F l^{-2}\left(1-\frac{2B^2}{\dil 1}\right)
\qquad &{\rm if}\ \dil 1\geq 2B^2\ ,\nonumber\\
4\Omega^2+\dil 1\geq 2B^2\qquad &{\rm if}\ \dil 1\leq 2B^2\ .
\nonumber
\end{subeqnarray}

\section{Connection with G\"odel-type metrics\label{sec4}}

\indent Solution \eref{genscalar}, which will 
be presented in this section, 
includes a number of known metrics as its 
particular cases. From set of the solutions of 
some physical interest (i.e. satisfying \eref{con} 
and the energy conditions) we present here such 
that recover G\"{o}del-type metrics \cite{ReboucasI}
\footnote{In \cite{ReboucasI} there is source-free 
electromagnetic field, which does not affect form 
of the metric.} in limiting process $C\rightarrow 0$. 

A transition from general \eref{genscalar} to important 
special cases can take either of the following three forms:

(i) With a choice $2P=-C^2\nu =\alpha^2>0,\ 
\gamma=2C\alpha^{-2}\sh^2\left(\frac{\alpha r}{2}\right)
\equiv {\tilde C}\sh^2\left(\frac{\alpha r}{2}\right)$, 
$\alpha$ is of inverse length dimension constant, 
one gets the metric
\begin{equation}
\begin{array}{lll}
 \d s^2 &=&\left[\d t+\frac{4\Omega}{\alpha^2}
\sh^2\left(\frac{\alpha r}{2}\right)\d
\varphi\right]^2-
\e^{2{\tilde C}\sh^2\left(\frac{\alpha r}{2}\right)}\d z^2\nonumber \\[2mm]
&-&\frac{2}{{\tilde C}^2\alpha^2}
\left[\e^{2{\tilde C}\sh^2\left
(\frac{\alpha r}{2}\right)}-2{\tilde C}(1-{\tilde C})
\sh^2\left(\frac{\alpha r}{2}\right)-1\right]\!
\d \varphi^2\nonumber\\[2mm]
 &-&\displaystyle\frac{{\tilde C}^2}{2}\frac{\sh^2(\alpha r)\e^{2{\tilde C}
\sh^2\left(\frac{\alpha r}
{2}\right)}\d r^2}
{\e^{2{\tilde C}\sh^2\left(\frac{\alpha r}{2}\right)}-2{\tilde C}
(1-{\tilde C})\sh^2\left(\frac {\alpha r}{2}\right)-1}\ ,
\label{sh}
\end{array}
\end{equation}
where $2\Omega^2=2B^2+\alpha^2\left(1-{\tilde C}\right)-\dil 1$, 
${\tilde C}\leq 0$ and $2B^2-{\tilde C}\alpha^2\geq \dil 1$.

(ii) Purely complex substitution $\alpha\rightarrow i\alpha$ 
in previous case gives the metric 
\begin{equation}
\begin{array}{lll}
 \d s^2 &=&\left[\d t+\frac{4\Omega}{\alpha^2}
\sin^2\left(\frac{\alpha r}{2}\right)\d
\varphi\right]^2
-\e^{2{\tilde C}\sin^2\left(\frac{\alpha r}{2}\right)}\d z^2\nonumber\\[2mm]
&-&\frac{2}{{\tilde C}^2\alpha^2}\left[1+2{\tilde C}
(1+{\tilde C})\sin^2\left(\frac {\alpha r}{2}\right)-
\e^{2{\tilde C}\sin^2\left(\frac{\alpha r}{2}\right)}\right]
\d \varphi^2\nonumber\\[2mm]
&-&\displaystyle\frac{{\tilde C}^2}{2}\frac{\sin^2(\alpha r)\e^{2{\tilde C}
\sin^2\left(\frac {\alpha r}{2}\right)}\d r^2}
{1+2{\tilde C}(1+{\tilde C})\sin^2\left(\frac {\alpha r}{2}\right)-
\e^{2{\tilde C}\sin^2\left(\frac{\alpha r}{2}\right)}}\ ,
\label{sin}
\end{array}
\end{equation}
together with $2\Omega^2=2B^2-\alpha^2\left(1+{\tilde C}\right)-\dil 1$, 
${\tilde C}\leq 0$ and let $2B^2-\alpha^2\geq\dil 1$.

(iii) Finally we pass to charged dust distribution, 
obtained from \eref{sh} or \eref{sin} 
when $\alpha$ tends to zero
\begin{equation}\label{alpha=0}
\d s^2=\left(\d t+\Omega r^2\d\varphi\right)^2-r^2\d\varphi^2-
\e^{Cr^2}(\d z^2+\d r^2)\ ,
\end{equation}
with $2\Omega^2=2B^2-2C-\dil 1$, $2(B^2-C)\geq \dil 1$ for 
$C\leq 0$, and $2(B^2-2C)\geq \dil 1$ for $0\leq C\leq \case {B^2}2$.

In each case \eref{sh}-\eref{alpha=0} the ranges 
of $C$ (or, equivalently, ${\tilde C}$) and $\phi_1$ 
are indicated for which the energy conditions in 
\tref{tab1} are fulfilled and the signature is correct. 

\begin{figure}[h]
\begin{center}\scalebox{0.8}{\includegraphics{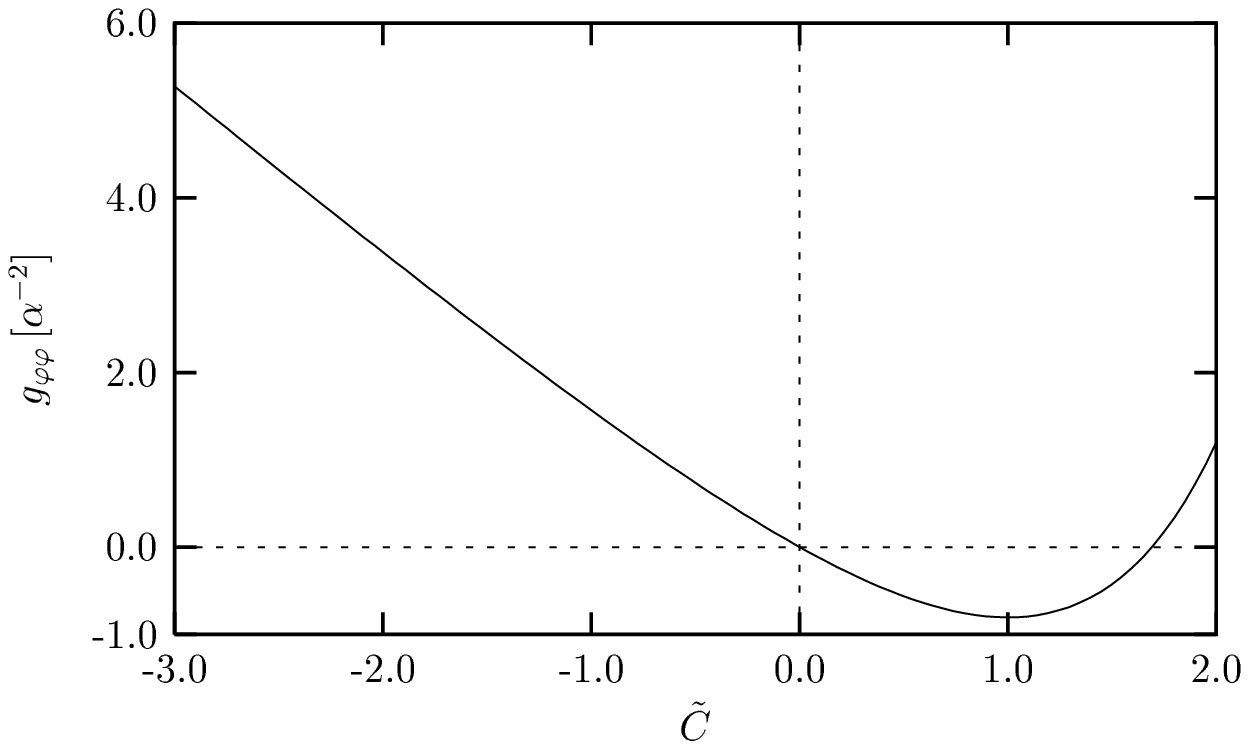}}\end{center}
\caption{For metric \eref{sin} some of the circles 
$t,\ z,\ r={\rm const}$ 
appear to be closed timelike curves e.g. if 
$r=r_n=(2n+1)\pi \alpha^{-1}$, $n$ is integer. 
On figure the values of $g_{\varphi\varphi}(r_n)$ 
(in units $\alpha^{-2}$) are plotted against 
${\tilde C}$ in extreme case $2B^2-\alpha^2=\dil 1$. 
Only ${\tilde C}\leq 0$ range is physically permissible.}
\label{fig1}\end{figure}

All of the above metrics \eref{sh}-\eref{alpha=0} admit 
closed timelike curves: \eref{sh} as long as 
$4\Omega^2 > \alpha^2$, \eref{sin} and \eref{alpha=0} for 
$\Omega\neq 0$ (see \fref{fig1} and \cite{{Klep},
{ReboucasI}} for details). Influence of the magnetic 
field to the chronology violation is positive unlike 
the scalar field influence, that is negative. 

\begin{figure}[h]
\begin{center}\scalebox{0.8}{\includegraphics{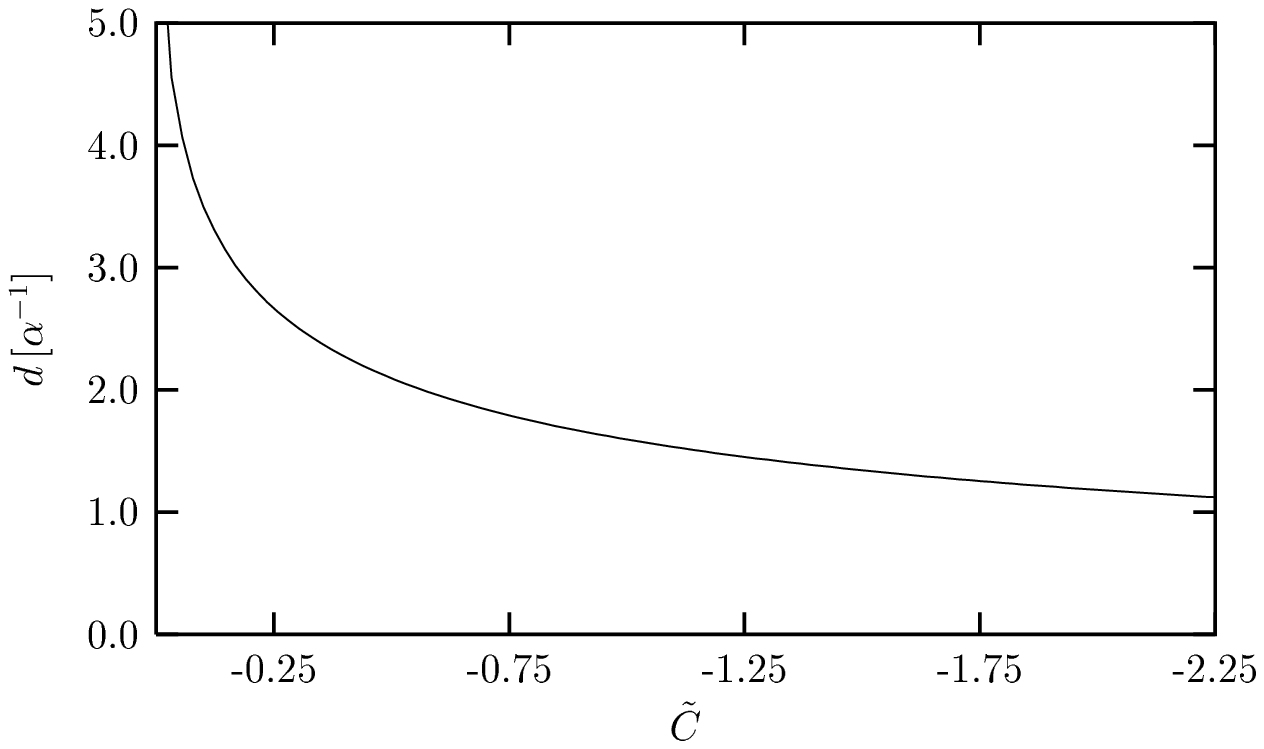}}
\end{center}
\caption{Smoothed dependence of proper radial distance 
$d$ (in units $\alpha^{-1}$) of infinity 
$r\rightarrow \infty$ 
from the rotation axis is plotted against ${\tilde C}$ 
for metric \eref{sh}.}\label{fig2}
\end{figure}

Metrics \eref{sh} or \eref{alpha=0} suffer from 
singularities for ${\tilde C}<0$ when $r$ goes to 
infinity, whereas Riemann tensor invariants of 
\eref{sin} are well behaved (see \cite{Klep}). Moreover  
\eref{sh} as well as \eref{alpha=0} are radially 
bounded in the sense that the proper radial 
distance $d$ of  infinity $r\rightarrow \infty$ 
from the rotation axis is finite unless ${\tilde C}=0$
(\fref{fig2}). For \eref{alpha=0} the last assertion 
is obvious and for \eref{sh} it follows since $d$ can 
be estimated as 
\[d=\int_0^\infty |\e^\delta|\d r\leq \alpha^{-1}
\int_0^\infty \frac{\e^{{\tilde C}x}}{\sqrt {x}}
\d x <\infty\ , \quad {\rm if}\ {\tilde C}<0\ .\]

\section{Concluding remarks\label{conclusion}}

A few points should be mentioned, before concluding 
this paper.  

As noted already before, metrics \eref{genscalar} 
and \eref{dust} contains arbitrary function as a 
consequence of the $r$-coordinate rescalling 
freedom. This arbitrarity can be locally 
removed by introducing a new coordinate $R$ defined 
by implicit equation $\d R=\e ^{\delta}\d r$.

To complete the picture, \eref{Wright} also 
provides us with solutions \eref{sh}-\eref{alpha=0} as 
its particular cases, namely with \eref{sh} subject to 
$2\Omega^2=(1-{\tilde C})\alpha^2$, furthermore with 
\eref{sin} for $2\Omega^2=-(1+{\tilde C})\alpha^2$, and 
with \eref{alpha=0} when $\Omega^2=-C$. 

\noindent Dust solution \eref{dust} for $\sigma=0$ 
contains \eref{alpha=0} with $\Omega^2=-C$. 

In section \ref{sec3} we have obtained the Wright 
solution alternative source. Thus by appropriate 
specialization of integration constants one can 
also obtain the G\"odel solution with alternative 
source formed by the combination of scalar field 
and charged fluid. It was realized in \cite
{ReboucasI} for $z$-dependent scalar field. 
Finally, we treat scalar field for the 
G\"odel metric having (in contrast to \cite
{ReboucasI}) non-trivial radial dependence. Let us 
take the G\"odel limit $C=0$ of \eref{sh} 
generated by \eref{Wright} into account.\footnote{%
  Metrics \eref{sin} and \eref{dust} do not 
  generate G\"odel-type solution besides the Minkowski 
  spacetime.}
In this case the scalar field 
can be written explicitly. Then equation \eref{dilaton} 
results in 
\begin{equation}\label{shdilaton}
\phi =\phi_0+\sqrt 2 Bz+\sqrt 2F_{\varphi z}
\ln \left|{\rm th}\frac{\alpha r}2\right|
\end{equation}
for the G\"odel limit of \eref{sh}.
The equation \eref{shdilaton} implies that $\phi$ 
diverges at $r=0$ and in $z$-infinities, which 
is consequence of equations \eref{rphi} and 
\eref{sing}. 

Note that the rest mass density \eref{Wrightdens} 
remains to be position dependent even after putting 
$C=0$ due to $F_{\varphi z}$ term contribution. In 
spite of this the total energy density is uniform 
through the spacetime, as it must be, insuring the 
spacetime homogeneity.

For further investigation of the charged scalar 
field and allowed frequencies in G\"odel-type 
background we refer the reader to \cite{Radu}, 
where is also considered the problem of field 
quatization using the Euclean 
approach to quantum field theory.

Let us summarize briefly the main results of the paper. 
We focused on the stationary cylindrically symmetric 
spacetime for which the matter content is constituted 
by the charged perfect fluid and massless scalar field. 
We dealt with the Lorentz force-free case only, with the 
magnetic field \eref{magnetic}. The fluid particles - 
carriers of the charge - rigidly rotate without acceleration. 
The solutions obtained are divided into two cases I and II.
For the case I the radial magnetic field is 
compensated by radial part of scalar field gradient, 
for the case II not. 

Case I consists of the Wright solution alternative source 
\eref{Wright} and the generalization \eref{genscalar} of 
the Lorentz force-free solutions in \cite{{Upornikov},{Klep}}. 
Case II describes charged dust solution \eref{dust} 
which is cylindrically symmetric spacetime 
but not regular at the origin. 

\section*{Acknowledgments}
The authors would like to express their acknowledgement 
to Dr. Unge and Prof. Lenc for helpful discussions. One of 
us (P K) also wishes to thank to D.~Ne\v{c}as, J.~Polcar 
and D. Hemzal for technical support.

\end{document}